\documentclass[onecolumn,A4paper,12pt]{article}

\usepackage{dcolumn}
\usepackage{bm}
\usepackage{times}
\usepackage[T1]{fontenc}
\usepackage[dvips]{graphicx}
\usepackage{graphics}
\usepackage{color}

\makeatletter

\newcommand{\erf}{{\rm erf}}

\long\def\symbolfootnote[#1]#2{\begingroup%
\def\thefootnote{\fnsymbol{footnote}}\footnote[#1]{#2}\endgroup}

\makeatother
\begin{document}

\begin{center}

{\large \bf On the analysis of composition profiles in binary diffusion couples: systems with a strong compositional dependence of the interdiffusion coefficient}

\vspace{0.2cm}
TAS KAVAKBASI Beng\"u$^{1,\rm a}$, GOLOVIN Igor S.$^{2,\rm b}$, PAUL Aloke$^{3,\rm c}$, DIVINSKI Sergiy V.$^{1,\rm d*}$

\vspace{0.2cm}
$^1$Institute of Materials Physics, University of M\"unster, Germany

\vspace{0.2cm}
$^2$Department of Physical Metallurgy of Non-Ferrous Metals, National University of Science and Technology 'MISiS', Moscow, Russia

\vspace{0.2cm}
$^3$Department of Materials Engineering, Indian Institute of Science, Bangalore, India

\vspace{0.2cm}

$1^{\rm a}$b\_task01@uni-muenster.de, $^{\rm b}$i.golovin@misis.ru, $^{\rm c}$aloke.paul@gmail.com, $^{\rm d}$divin@uni-muenster.de

\end{center}


\vspace{0.3cm}
\noindent {\bf Abstract.} Diffusion couple technique is an efficient tool for the estimating the chemical diffusion coefficients. Typical experimental uncertainties of the composition profile measurements complicate a correct determination of the interdiffusion coefficients via the standard Boltzmann-Matano, Sauer-Freise or the den Broeder methods, especially for systems with a strong compositional dependence of the interdiffusion coefficient. A new approach for reliable fitting of the experimental profiles with an improved behavior at both ends of the diffusion couple is proposed and tested against the experimental data on chemical diffusion in the system Fe--Ga.


\subsection*{Introduction}

\noindent Diffusion couple technique is widely used to address chemical diffusion between the given end-members in binaries as well as in multi-component systems \cite{book}. A composition-dependent interdiffusion coefficient, $\tilde{D}(C)$, can be determined from a measured single concentration profile, $C(x)$, directly in binary and pseudo-binary multicomponent systems (here $x$ is the coordinate and $C$ is the concentration in [mol/m$^3$]). The experimental methods for estimating the composition-dependent interdiffusion coefficients in multicomponent systems are far more complex. It needs several samples to estimate them in a ternary system and simply not possible in a system with a higher number of components (unless a pseudo-binary \cite{Aloke} approach is followed).  Therefore, our discussion in this article is mainly on the binary systems. However, a similar method can be followed in multicomponent systems  for estimation of the interdiffusion flux \cite{book}.

One of the most frequently methods is the Boltzmann-Matano \cite{Matano, Boltz} approach. In that follows we will use the atomic fractions $N$ instead of the concentrations $C$, which are related as $C=N/V_m$, with $V_m$ being the molar volume. If $V_m$ is constant or varies nearly ideally (following Vegard's law) with the composition, one obtains \cite{book},

\begin{equation}
 \tilde{D}(N^*) = - \frac{1}{2t} \left( \left. \frac{dN}{dx}\right|_{N^*} \right)^{-1} \int \limits_{N^-}^{N^*}(x-x_M) dN. \label{eq:BM}
\end{equation}

\noindent Here $N^-$ is the concentrations at the left-hand-side end of the diffusion couple (correspondingly, $N^+$ would be that at its right-hand-side end) and $N^*$ is the given concentration from the interval under consideration, $N^- \le N^* \le N^+$. The distance is measured from the position of the Matano plane, $x=x_M$, which is defined as a plane where the flux of mass from the left to the right and from the right to the left ends is balanced,

\begin{equation}
\int \limits_{N^-}^{N^M}x dN = \int \limits_{N^M}^{N^+}x dN. \label{eq:matano}
\end{equation}

\noindent Here $N^M$ is the concentration at $x=x_M$. Note that the method can further be elaborated for multi-component systems, too, within the so-called pseudo-binary approach \cite{Aloke, Aloke2}. 

\begin{figure}[t]  
\begin{minipage}[b]{0.5\linewidth}
\includegraphics[width=7cm]{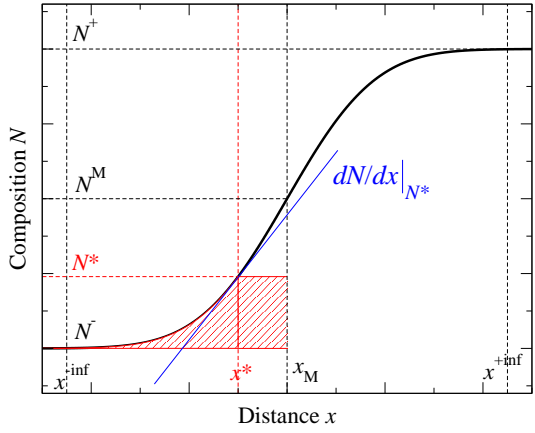} 
\end{minipage}
\hfil
\begin{minipage}[b]{0.5\linewidth}
\caption{\label{fig:sketch}\small Schematic determination of the interdiffusion coefficient according to Boltzmann-Matano method \cite{Matano,Boltz}, see Eq.~\ref{eq:BM}. A concentration profile $N(x)$ is measured in the diffusion couple with $N^-$ and $N^+$ end-concentrations and $\tilde{D}$ is evaluated for the concentration $N^*$ corresponding to the position $x^*$. $x_M$ is the position of the Matano plane.}
\end{minipage}
\end{figure}

The determination of the interdiffusion coefficient is sketched in Fig.~\ref{fig:sketch}.

The analysis will be simplified if the relative concentration, $Y = \frac{N-N^-}{N^+-N^-}$, is introduced as in the method of Sauer-Freise \cite{SF} or de Broeder \cite{dB}, and the interdiffusion coefficient will be determined as \cite{book},

\begin{equation} 
 \tilde{D}(C^*) = \frac{1}{2t} \left( \left. \frac{dY}{dx}\right|_{Y^*} \right)^{-1}
 \left[ (1-Y^*) \int \limits_{x^-}^{x^*}Y dx + Y^* \int \limits_{x^*}^{x^+}(1-Y) dx \right]  \label{eq:SF}
\end{equation}

\noindent where $Y^* = (N^*-N^-)/(N^+-N^-)$.

In fact, the interdiffusion coefficient is proportional to the ratio of the integral over the concentration profile (the hatched area in Fig.~\ref{eq:BM}) to its first derivative (drawn as a line in Fig.~\ref{eq:BM}, too). This means that the integral should go to zero equally fast as the corresponding derivative, $dN/dx$, as $x^* \rightarrow x^{-}$ (or equivalently at $x^* \rightarrow x^{+}$), in order to provide reasonable estimates of $\tilde{D}$ at both extremes. 

Typically, the concentration profile is measured using scanning electron microscopy (SEM) equipped with an EDX detector (accuracy of the concentration determination of about 0.5 to 1\%) or electron probe microanalysis (EPMA) with a WDX detector and significantly improved accuracy (about 0.1 to 0.2 \%). In any case, due to inherent uncertainties of the concentration determination, the individual data points scatter considerably and a fitting procedure is applied. Basically, if the concentration profile is fitted by a polynomial function (typically piece-wise) \cite{Kapoor}, it is easy to show that  $\tilde{D} \rightarrow 0$ as $x^* \rightarrow x^{\pm}$. 

Indeed, approximating $N(x)$ by a polynomial function at $x^* \rightarrow x^{-}$, $N(x)=N^- + b \cdot (x-x^-)^n + ...$, and keeping only the term with the lowest power $n\ge1$, the value of the integral in Eq.~(\ref{eq:BM}) is estimated as (the total area of the hatched region in Fig.~\ref{fig:sketch})

\begin{equation}
 integral \simeq \frac{b}{(n+1)} (x-x^-)^{n+1} + b(x-x^-)^n x^-
\end{equation}

\noindent and the derivative is simply $nb(x-x^-)^{n-1}$ (the higher-order terms are neglected). Substituting these expressions in Eq.~(\ref{eq:BM}) one sees immediately that $\tilde{D} \propto (x-x^-)$ and  $\tilde{D} \rightarrow 0$ as $x^* \rightarrow x^{-}$.

A most suitable approach is the usage of $\erf$-function(s) for profile fitting since this method provides inherently a more reasonable behavior at both ends of the diffusion couple, since the erf-function corresponds to an {\it exact} solution of the diffusion problem for a particular case of the constant diffusion coefficient \cite{book}.

The present paper aims to examine applicability of different fits of the concentration profiles which would provide best results for determination of the interdiffusion coefficients. Experimentally, interdiffusion in the Fe--Ga alloys is investigated applying a diffusion couple technique. 

\subsection*{Experimental Details}


\noindent FeGa alloys of two compositions, Fe--8at.\%Ga and Fe--24at.\%Ga were prepared by casting. The ingots were cut into rectangular shape of $6 \times 6 \times 1$~mm$^3$ in size. Both sample surfaces were polished to a mirror-like quality. 

The two end-members were fixed together in a custom-built fixture produced from heat-resistant steel (THERMAX). To prevent any contact of the samples and fixture, the samples were wrapped in tantalum foil. The fixture was sealed under argon atmosphere in a quartz tube and annealed at 1173~K for 5~hours. According to the phase diagram \cite{Mas}, this temperature corresponds to a single-phase disordered solid solution of Ga atoms in $\alpha$-Fe lattice (the A2 structure) for all compositions between the given end-members.

The chemical profile was measured by using a JEOL JXA 8900 Superprobe Electron Microprobe Microanalysis System (EPMA) in the group of Dr. Jasper Berndt-Gerdes (Institute of Mineralogy, University of M\"unster, Germany). A cross-section of the diffusion couple was produced by cutting of the samples mounted in an epofix resin and subsequent polishing. A line scan analysis was done to the distances of $500~\mu$m from the interface on both sides of the couple with a step size of $2~\mu$m.

\subsection*{Results}

\begin{figure}[ht]  

\begin{minipage}[b]{0.45\linewidth}
\includegraphics[width=7cm]{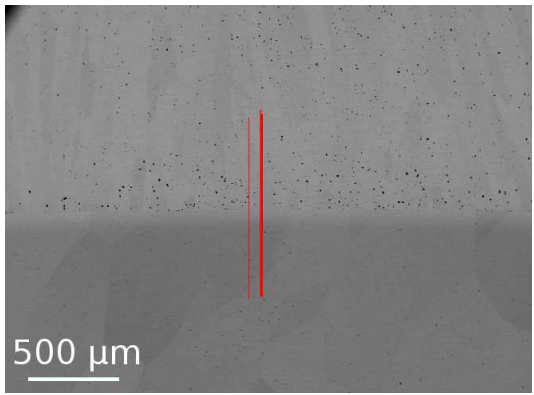}  
\end{minipage}
\hfill
\begin{minipage}[b]{0.5\linewidth}
\caption{\label{fig:SEM} \small A SEM image of the cross-section of the Fe--Ga diffusion couple. The red lines indicate location of the line scans with EPMA.}
\end{minipage}
\end{figure}

\begin{figure*}[t]
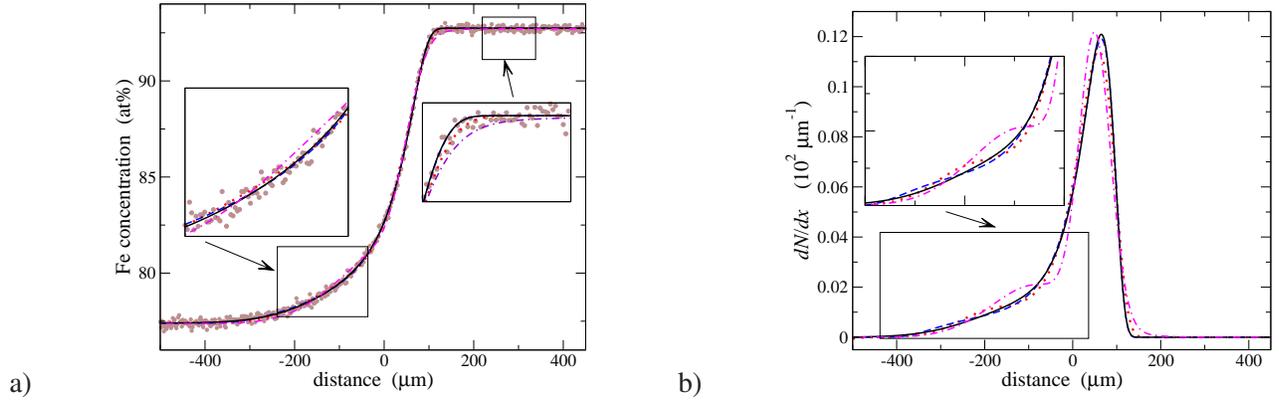
  
\begin{center}
\small
a)\hspace{1cm} \includegraphics[width=6.3cm]{chem-prof.eps}
\hspace{1cm}
b)\hspace{1cm} \includegraphics[width=6.5cm]{chem-der.eps}
\end{center}
\caption{\label{fig:fit} \small Measured chemical diffusion profile and its fit by several model functions (a) and the corresponding derivatives (b). 
The profile is fitted according to Eq.~(\ref{eq:sum}) (magenta dashed-dotted line), Eq.~(\ref{eq:poly}) (red dotted line), Eq.~(\ref{eq:ipoly}) (blue dashed line) and Eqs.~(\ref{eq:my})--(\ref{eq:my2}) (black solid line). The insets zoom the given regions.
}
\end{figure*}

\noindent
A SEM image of the cross-section is shown in Fig.~\ref{fig:SEM}, with red lines indicating the positions of line-scans. The measured concentrations were converted to at.\% and the resulting  chemical profile (i.e., Fe concentration in at.\%) is given in Fig.~\ref{fig:fit}a. The position of the Matano plane was determined according to Eq.~(\ref{eq:matano}) and the origin of the abscissa is set to $x=x_M$.

A strong deviation of the element distribution from simple error function solution is obvious. A very strong compositional dependence of the interdiffusion coefficient is pronounced, too.

\subsection*{Discussion}

\noindent One of standard approaches to fit such profile is the usage of a sum of complementary error functions \cite{Zhang} (as a sum of $m$ $\erf$-functions),

\begin{equation}\label{eq:sum}
N(x,\mu) = N^- + \sum_{i=1}^m A_i \left[1+ \erf \left( \frac{x-x_i}{\mu_i} \right) \right] 
\end{equation}

\noindent with $A_i$, $x_i$ and $\mu_i$ as the fit parameters (fixing the concentration of Fe atoms at the left end of the couple, $N^-$). In following we will fix the number of fit parameters to 5 for all fit functions and, thus, will keep only two terms in the sum above ($m=2$). Note that in this case only one parameter, i.e. $A_1$ or $A_2$, will be a free parameter in view of the obvious relation

\begin{equation}
 N^+ - N^- = 2A_1 + 2A_2. \label{eq:bal}
\end{equation}

We have chosen $A_2$ as a free parameter and evaluated $A_1$ via Eq.~(\ref{eq:bal}). A fit according to Eq.~(\ref{eq:sum}) is shown in Fig.~\ref{fig:fit}a by magenta dashed-dotted line and the determined parameters are listed in Table~\ref{tab:data}. While the composition profile is fitted well (the fit quality will definitely be improved if three or even more terms will be used in  Eq.~(\ref{eq:sum})), the corresponding derivative, $dN/dx$, reveals specific step-like artifacts, see Fig.~\ref{fig:fit}b, which will affect the subsequent determination of the interdiffusion coefficient (see below). 

A similar behavior of the corresponding derivatives, $dN/dx$, would be observed if a sum of Boltzmann functions \cite{Chen, Li} would be used for the fitting of the concentration profile, i.e.

\begin{equation}\label{eq:sumB}
N(x,\mu) = N^- + \sum_{i=1}^m \frac{A_i}{1+ \exp \left( \frac{x-x_i}{\mu_i} \right)}  
\end{equation}

Again, characteristic 'steps' would be present in $dN/dx$ and correspondingly in the determined interdiffusion coefficient.

One may argue that a most reliable way is a usage of an erf-type of solution and fitting of the arguments by a suitable function. This approach is based on the fact that Eq.~(\ref{eq:sum}) with $m=1$ is exact solution in the case of a constant interdiffusion coefficient. Thus, using Eq.~(\ref{eq:sum}) with $m=1$ and choosing an appropriate functional form for its argument would mimic the compositional dependence of the interdiffusion coefficient.

A simplest approach would be just a polynomial fit as

\begin{equation}\label{eq:poly} 
 N(x,\mu) = N^- + \frac{N^+-N^-}{2} \times 
 \left[1 + \erf \left\{ (x-x_0)(A_0+A_1 x + A_2 x^2 + A_3 x^3) \right\} \right]
\end{equation}

\noindent Note that for consistency, we are keeping the number of independent fit parameters equal to five.

The best fit is drawn in Fig.~\ref{fig:fit}a as red dotted line and a generally good fit is obtained. There are some problems at the distances far from the position of the Matano plane which would be inherent for polynomial fit as in Eq.~(\ref{eq:poly}), but they are not critical for the subsequent determination of the interdiffusion coefficient since one may expect large deviations at the concentrations corresponding to the end-members. Some systematic deviations from the experimental points at distances about $+100~\mu$m are due to our limitation to fourth-order terms in the polynomial expression above.

Since the interdiffusion coefficient enters the parameter $\mu$ in Eq.~(\ref{eq:sum}), a good idea would be a usage of an inverse polynomial function like

\begin{equation}\label{eq:ipoly}
N(x,\mu) = N^- + \frac{N^+-N^-}{2} \times
\left[1 + \erf \left\{ \frac{x-x_0}{A_0+A_1 x + A_2 x^2 + A_3 x^3} \right\} \right]
\end{equation}

The corresponding fit is shown in Fig.~\ref{fig:fit}a by blue dashed line and it is almost not distinguishable from the black solid line. Some deviations are seen only for the corresponding derivatives, Fig.~\ref{fig:fit}b, especially at distances of about $-200~\mu$m. 

The main problem of the functional form of Eq.~(\ref{eq:ipoly}) is the existence of poles for the inverse polynomial function that hinders a straightforward fitting procedure. A very careful selection of the initial guesses is required.

Various tests have shown that the following expression is most suitable for a robust and reliable fit of the concentration profiles in the present study,

\begin{equation}\label{eq:my}
N(x,\mu) = N^- + \frac{N^+-N^-}{2} \times \left[1+ \erf \left( \frac{x-x_0}{\mu} \right) \right] 
\end{equation}

\noindent with the parameter $\mu$ determined as

\begin{equation}\label{eq:my2}
\mu = \mu_0 + \mu_1 \times \left[1+ \erf \left( \frac{x-x_{\mu}}{\sigma} \right) \right] 
\end{equation}

The corresponding fit is shown in Fig.~\ref{fig:fit}a by black solid line and the corresponding derivative, Fig.~\ref{fig:fit}b, reveals less 'features' which may be considered as artifacts of the fitting procedure.

Table~\ref{tab:data} summarize all fit parameters determined in the present study for the experimentally determined concentration profile.

\begin{figure}[ht]  
\begin{minipage}[b]{0.5\linewidth}
\includegraphics[width=7cm]{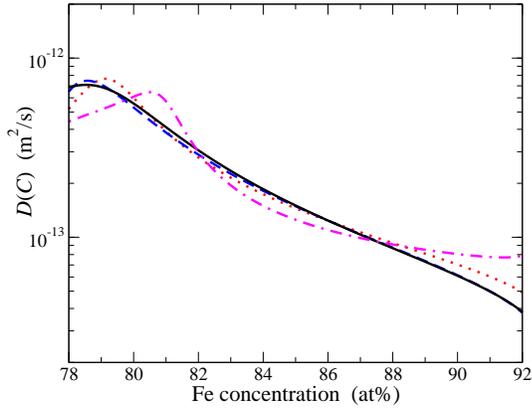}
\end{minipage}
\hfill
\begin{minipage}[b]{0.5\linewidth}
\caption{\label{fig:dif} \small The composition dependence of the interdiffusion coefficient $\tilde{D}(C)$ for all models considered in the paper: Eq.~(\ref{eq:sum}) (magenta dashed-dotted line), Eq.~(\ref{eq:poly}) (red dotted line), Eq.~(\ref{eq:ipoly}) (blue dashed line), and Eqs.~(\ref{eq:my}) \& (\ref{eq:my2}) (black solid line).}
\end{minipage}
\end{figure}

In Fig.~\ref{fig:dif} the results of determination of the interdiffusion coefficients as function of the composition using the Sauer-Freise-de Broeder approach \cite{SF, dB}, Eq.~(\ref{eq:SF}), are shown for all fit functions, Eqs.~(\ref{eq:sum}), (\ref{eq:poly}), (\ref{eq:ipoly}), and (\ref{eq:my}) \& (\ref{eq:my2}).

\begin{table*}[t]
\begin{center}
\caption{\small The numerical values of the fit parameters for best fits using different approaches considered in the paper. The distance is measured in microns.}
\small
\label{tab:data}
\vspace{0.3cm}
\begin{tabular}{ccccc}
\hline
\multicolumn{5}{c}{model Eq.~(\ref{eq:sum})} \\
$x_1$    & $\mu_1$ & $A_2$    &$x_2$ & $\mu_2$\\ 
$-80.89$ & $145.3$ & $0.6477$ & $50.70$ & $49.78$ \\ \hline
\multicolumn{5}{c}{model Eq.~(\ref{eq:poly})} \\
$x_0$    & $A_0$ & $A_1$    &$A_2$&$A_3$\\ 
$31.43$  & $0.930 \times 10^{-2}$ & $5.039 \times 10^{-5}$ & $1.844 \times 10^{-7}$ & $0.209 \times 10^{-9}$ \\ \hline
\multicolumn{5}{c}{model Eq.~(\ref{eq:ipoly})} \\
$x_0$    & $A_0$ & $A_1$    &$A_2$&$A_3$\\ 
$32.15$  & $112.70$ & $-0.588$ & $-0.894 \times 10^{-4}$ & $2.688 \times 10^{-6}$ \\ \hline
\multicolumn{5}{c}{model Eqs.~(\ref{eq:my}) \& (\ref{eq:my2})} \\
$x_0$    & $\mu_0$ & $\mu_1$    &$x_{\mu}$&$\sigma$\\ 
$32.55$  & $214.6$ & $-105.9$ & $7.962$ & $189.9$ \\
\hline
\end{tabular}
\end{center}
\end{table*}

An analysis of the data presented in Fig.~\ref{fig:dif} allows to draw several important conclusions.

\begin{itemize}
 \item A usage of a sum of erf-functions for fitting of the concentration profiles, Eq.~(\ref{eq:sum}), is simple, however, one has to evaluate carefully all 'features', i.e. ghost extremes, for the derived interdiffusion coefficients.
 \item A polynomial expression, Eq.~(\ref{eq:poly}), could be considered as suitable, however, it provides reasonable estimates of the interdiffusion coefficients in a smaller interval of the concentrations.
 \item An inverse polynomial expression, Eq.~(\ref{eq:ipoly}), provides very good fits, however, it is prone to fit problems via appearance of poles. A very careful choice of the initial values of the fit parameters is required.
 \item An exponential fit, Eqs.~(\ref{eq:my}) \& (\ref{eq:my2}), or a power law one (like $\mu = \mu_0 \times N^m$ instead of Eq.~(\ref{eq:my2})) provide most robust and reliable fit and straightforward determination of the interdiffusion coefficients.
\end{itemize}

\subsection*{Summary}

\noindent In this paper different functions are tested for fitting the experimentally determined concentration profile in the Fe--Ga system with a strong compositional dependence of the interdiffusion coefficient. 

A most reasonable results are obtained using functional fitting of the argument of an erf-function. A specific expression, Eqs.(\ref{eq:my})--(\ref{eq:my2}), is proposed which mimic the compositional dependence of the interdiffusion coefficient and which provide the best results. 

This expression is proposed to use for fitting of the composition profiles in binary as well as in multicomponent systems. An extension of the approach to possible up-hill diffusion would be straightforward.

\vspace{0.3cm}
\noindent{\bf Acknowledgments.} The authors thank to Dr. Jasper Berndt-Gerdes (Institute of Mineralogy, University of M\"unster, Germany) for a help with EPMA measurements. The financial support of the Deutsche Forschungsgemeinschaft through the research grant (Di 1419/11-1) is acknowledged.

\renewcommand{\refname}

\end{document}